\documentclass[10pt,aps,prd,showpacs,superscriptaddress,twocolumn,nofootinbib,hidelinks]{revtex4-2}
\usepackage{hyperref}
\usepackage{verbatim} 
\usepackage{amsfonts}
\usepackage{amssymb}
\usepackage{amsmath}
\usepackage{MnSymbol}
\usepackage{graphicx}
\usepackage[caption=false]{subfig}
\usepackage{enumitem}
\usepackage{xcolor}
\usepackage{appendix}

\date{\today}

\begin{document}

\title{A relativistic quantum broadcast channel}
\author{Ian Bernardes Barcellos}\email{ian.barcellos@ufabc.edu.br}

\author{Andr\'e G.\ S.\ Landulfo}\email{andre.landulfo@ufabc.edu.br}

\affiliation{Centro de Ci\^encias Naturais e Humanas,
Universidade Federal do ABC, 
Avenida dos Estados, 5001, 09210-580, Bangu,
Santo Andr\'e, S\~ao Paulo, Brazil}

\begin{abstract}

We investigate the transmission of classical and quantum information between three observers in a general globally hyperbolic spacetime using a quantum scalar field as a communication channel. We build a model for a quantum broadcast channel in which one observer (sender) wishes to transmit (classical and quantum) information to two other observers (receivers). They possess some localized two-level quantum system (a qubit) that can interact with the quantum field in order to prepare an input or receive the output of this channel. The field is supposed to be in an arbitrary quasifree state, the three observers may be in arbitrary states of motion, and no choice of representation of the field canonical commutation relations is made. The interaction of the field and qubits is such that it allows us to obtain the map that describes this channel in a non-perturbative manner. We conclude by analyzing the rates at which information can be transmitted through this channel and by investigating relativistic causality effects on such rates.

\end{abstract}

\maketitle

\section{Introduction}
\label{sec:introduction}

Network information theory is the area of knowledge that studies classical communication problems involving multiple parts. Here,  the word ``classical'' stands not only for the fact that the information being transmitted is classic (bits) but also for the physical systems in which such information is encoded, i.e., systems that can be described by some area of classical physics (such as Electromagnetism). One particular case of interest is the broadcast channel, where typically one sender wishes to transmit information to multiple receivers (like radio and TV stations broadcasting their signals, for example).

Nowadays, one of the main goals of quantum information theory is to extend several results of information theory to the quantum world~\cite{nielsen_chuang-2010,wilde-2013}, investigating any new features or advantages that can arise when one uses quantum systems to encode, process, and transmit information. The quantum network information theory comprises the studies of communication protocols using quantum systems to convey classical (bits) or quantum (qubits) information. In particular, the classical broadcast channels can be extended to the so-called \textit{quantum broadcast channels}, where one sender transmits classical or quantum input information to many receivers using a quantum system as a communication channel with quantum outputs~\cite{SW2015, YHD2011, DHL2010, WDW17}.

Such communication scenarios are very suitable for analyzing how relativistic effects can influence one's ability to communicate using quantum channels. This could be due to the existence of nontrivial spacetime structures such as black hole event horizons, Cauchy horizons, and causal horizons arising from the relativistic relative motion between senders and receivers or even due to the expansion of spacetime~\cite{wald-1984}.

In order to consistently analyze quantum information theory in general spacetimes, one should use quantum field theory in curved spacetimes (QFTCS)~\cite{wald-1994}. This approach was used by several authors to analyze the communication process in relativistic settings, with particular attention being paid to Minkowski~\cite{AM03, BHTW10, LT13, MHM12, BHP12, CK10, JMK14, MM15, JRMK18, J17, HLL12, SAKM20,YASKM20}, Schwarzschild~\cite{HBK12, BA14, BA15, JACKM20, QW17}, or asymptotically flat cosmological spacetimes~\cite{MM, BGMM16,SM17}. However, only recently~\cite{landulfo-2016} a communication model valid in general globally hyperbolic spacetimes and in which the parts that convey information can move in arbitrary worldlines and interact with the quantum field (used as communication channel)  only in the vicinity of its worldlines was developed (and, since then, other works in this context have emerged as, for instance, Refs.~\cite{TY22, LMBM23}). This is interesting for two reasons: firstly, it allows the analysis of information exchange between more general observers, not only observers following orbits of some Killing field (which does not even exist in spacetimes lacking timelike symmetries). Secondly, the model studied in~\cite{landulfo-2016} allows one to investigate the outputs of the quantum communication in a nonperturbative manner and thereby is suitable to investigate both the causality as well as the communication between parts lying in early and future asymptotic regions (limits that would invalidate results obtained by perturbative methods).

In the present paper, we generalize the analysis of~\cite{landulfo-2016}. This is done by constructing a model for classical-quantum, quantum-quantum, and entanglement-assisted quantum-quantum broadcast channels. We consider an arbitrary globally hyperbolic spacetime in which one observer (Alice) wants to convey classical (or quantum) information to two receivers (Bob and Charlie) using a quantum scalar field as a communication channel. The three observers will use two-level quantum systems (qubits) to locally interact with the quantum field in order to send or receive information. The observers may be in arbitrary states of motion, the interaction between the detectors and the field is similar to the one given by the Unruh-DeWitt model~\cite{dewitt-1979}, and the field may initially be in an arbitrary quasifree state~\cite{wald-1994}. We suppose, however, that the two levels of each qubit have the same energy. This model is interesting because the evolution of the system can be computed exactly, and therefore we will obtain nonperturbative results for the communication rates associated with such a broadcast channel. As we will see, causality in the information exchange is explicitly manifest in our results.

This work is organized as follows. In Sec.~\ref{sec:field-quantization} we will present the quantization procedure of a free scalar field on a globally hyperbolic spacetime as well as the class of states we will be using. In Sec.~\ref{sec:broadcast-channel} we describe the interaction between the qubits and the field and determine the quantum map that relates the information Alice wants to convey to the final joint state of Bob's and Charlie's qubits. In Sec.~\ref{sec:achievable-rates} we investigate the rates at which information can be transmitted using this broadcast channel, as well as the influence of the spacetime curvature or relative motion of observers in the communication process. In Sec.~\ref{sec:conclusions} we give our final remarks. We assume metric signature $(-+++)$ and natural units in which $c=\hbar=G=k_B=1$, unless stated otherwise.

\section{Field Quantization}
\label{sec:field-quantization}

Let us consider a free, real scalar field $\phi$ propagating in an arbitrary four-dimensional globally hyperbolic spacetime $\left(\mathcal{M},g\right)$, where $\mathcal{M}$ denotes the four-dimensional spacetime manifold and $g$ its Lorentzian metric. Let the spacetime be foliated by Cauchy surfaces $\Sigma_t$ labeled by the real parameter $t$. The field is described by the action
\begin{equation}
  \label{eq:KG-action}
  S\equiv- \frac{1}{2}\int_\mathcal{M}\epsilon_\mathcal{M}\,
  ( \nabla_a\phi \nabla^a\phi + m^2 \phi^2 + \xi R \phi^2 ),
\end{equation}
where $\epsilon_\mathcal{M}=\sqrt{-\mathfrak{g}}dx^0\wedge \cdots \wedge dx^3$ is the spacetime volume 4-form, $m$ is the field mass, $\xi\in\mathbb{R}$, $R$ is the scalar curvature, $\nabla_a$ is the torsion-free covariant derivative compatible with the metric $g$, and $\mathfrak{g}\equiv\text{det}(g_{\mu\nu})$ in some arbitrary coordinate system $\{x^\mu\}$. The extremization of the  action~\eqref{eq:KG-action} gives rise to the Klein-Gordon equation  
\begin{equation}
  \label{eq:KG-equation}
  (-\nabla^a\nabla_a + m^2+\xi R)\phi = 0.
\end{equation}

In the canonical quantization procedure, we promote the real field $\phi$ to an operator\footnote{Rigorously, an operator-valued distribution.} that satisfies the ``equal-time'' canonical commutation relations (CCR)
\begin{equation}
\label{eq:CCR-unsmeared-1}
[\phi (t, {\bf x}), \phi (t, {\bf x}') ]_{\Sigma_t} =
[\pi  (t, {\bf x}), \pi  (t, {\bf x}') ]_{\Sigma_t} = 0,
\end{equation}
\begin{equation}
\label{eq:CCR-unsmeared-2} 
[ \phi (t, {\bf x}), \pi  (t, {\bf x}')]_{\Sigma_t} =
i \delta^3 ({\bf x}, {\bf x}' ),
\end{equation}
where $\mathbf{x}\equiv(x^1,x^2,x^3)$ are spatial coordinates on $\Sigma_t$ and $\pi(x)$ is the conjugate momentum defined as
\begin{equation}
  \label{eq:conjugate-momentum-definition}
  \pi\equiv\frac{\delta S}{\delta\dot{\phi}}\,,
\end{equation}
with the notation $\dot{\phi}\equiv\partial_t\phi$. In addition, we may formally write the canonical Hamiltonian of the field as
\begin{equation}
  \label{eq:field-canonical-hamiltonian}
  H_\phi(t)\equiv\int_{\Sigma_t}d^3{\bf x} \,\left(\pi(t,\mathbf{x})\dot\phi(t,\mathbf{x})-\mathcal{L}[\phi,\nabla_a\phi]\right),
\end{equation}
with
\begin{equation}
     d^3{\bf x}\equiv dx^1\wedge d x^2 \wedge dx^3
\end{equation}
and
\begin{equation}
  \label{eq:field-lagrangian-density}
  \mathcal{L}[\phi,\nabla_a\phi]\equiv-\frac{1}{2}\sqrt{-\mathfrak{g}} \,( \nabla_a\phi \nabla^a\phi + {m}^2 \phi^2 + \xi R \phi^2)
\end{equation}
being the Lagrangian density.

To find a representation of the CCR, Eqs.~\eqref{eq:CCR-unsmeared-1} and~\eqref{eq:CCR-unsmeared-2}, we define an antisymmetric bilinear map $\sigma$ acting on the space $\mathcal{S}^\mathbb{C}$ of complex solutions of Eq.~\eqref{eq:KG-equation} as
\begin{equation}
  \label{eq:antisymmetric-bilinear-map}
  \sigma(\psi_1,\psi_2)\equiv\int_{\Sigma_t}\epsilon_\Sigma  \,n^a\left[\psi_2\nabla_a\psi_1-\psi_1\nabla_a\psi_2\right],
\end{equation}
where $\epsilon_\Sigma$ represents the proper-volume 3-form on the Cauchy surface $\Sigma_t$ and $n^a$ its future-directed normal unit vector. It allows us to define the Klein-Gordon product as
\begin{equation}
  \label{eq:KG-inner-product}
  \langle\psi_1,\psi_2\rangle\equiv -i\,\sigma(\overline{\psi}_1,\psi_2),
\end{equation}
and, although this product is not positive-definite on $\mathcal{S}^\mathbb{C}$, we may choose any subspace $\mathcal{H}\subset\mathcal{S}^\mathbb{C}$ (the so-called \textit{one-particle Hilbert space)} such that: \textbf{(i)}~$\mathcal{S}^\mathbb{C}\simeq\mathcal{H}\bigoplus\overline{\mathcal{H}}$;\footnote{For the sake of mathematical precision, we note that one must first suitably Cauchy-complete  $\mathcal{S}^\mathbb{C}$ for this decomposition to be valid.} \textbf{(ii)}~the KG product is positive definite on $\mathcal{H}$, thus making $(\mathcal{H},\langle,\rangle)$ a Hilbert space;\footnote{After its completion with respect to the norm induced by $\langle,\rangle$.} \textbf{(iii)}~given any $u\in\mathcal{H}$ and $v\in\overline{\mathcal{H}}$, $\langle u,v\rangle=0$. (See~\cite{wald-1994} for details.) The Hilbert space that comprises the field states is defined as the symmetric Fock space $\mathfrak{F}_s(\mathcal{H})$ and the quantum field operator is formally defined as
\begin{equation}
  \label{eq:unsmeared-field-operator}
  \phi(t,\mathbf{x})\equiv\sum_j\left[u_j(t,\mathbf{x})a(\overline{u}_j)+\overline{u}_j(t,\mathbf{x})a^\dagger(u_j)\right],
\end{equation}
where $\{u_j\}$ comprise an orthonormal basis for $\mathcal{H}$ and $a(\overline{u})$/$a^\dagger(v)$ are the usual annihilation/creation operators associated with the modes $u$/$v$, respectively. They satisfy the commutation relations
\begin{equation}
  \label{eq:commutation-relation-annihilation-and-creation}
  \left[a(\overline{u}),a^\dagger(v)\right]=\langle u,v\rangle I,
\end{equation}
with $I$ being the identity operator on $\mathfrak{F}_s(\mathcal{H})$. The vacuum state associated with this representation of the CCR is the normalized vector $|0\rangle$ that satisfies $a(\overline{u})|0\rangle=0$ for every mode $u\in\mathcal{H}$.

In order to make it mathematically well-defined, the quantum field operator must be defined as an operator-valued distribution. To this end, let  $\mathcal{S}\subset\mathcal{S}^\mathbb{C}$ be the space of real solutions of Eq.~\eqref{eq:KG-equation} whose restriction to Cauchy surfaces have compact support and $K:\mathcal{S}\rightarrow\mathcal{H}$ be the projection operator that takes the positive-norm part of any $\psi\in\mathcal{S}$. If $C^\infty_0(\mathcal{M})$ denote the set of all smooth compactly-supported real functions on $\mathcal{M}$, we define the map $E:C^\infty_0(\mathcal{M})\rightarrow\mathcal{S}$ acting on some \textit{test function}  $f\in C^\infty_0(\mathcal{M})$ as
\begin{equation}
  \label{eq:def-causal-propagator}
  Ef(x)\equiv Af(x)-Rf(x),
\end{equation}
where $Af$ and $Rf$ are the advanced and retarded solutions of the Klein-Gordon equation with source $f$, respectively. Hence, they satisfy 
\begin{equation}
  \label{eq:KG-equation-source}
  P(Af) = P(Rf) = f,
\end{equation}
with $P\equiv-\nabla^a\nabla_a + m^2+\xi R$ representing the Klein-Gordon differential operator. 

Now, for each test function $f\in C^\infty_0(\mathcal{M})$, we define a \textit{smeared quantum field operator} by
\begin{equation}
  \label{eq:smeared-quantum-field-definition}
  \phi(f)\equiv i\left[a(\overline{KEf})-a^\dagger(KEf)\right],
\end{equation}
which satisfies the covariant version of the CCR,
\begin{equation}
  \label{eq:covariant-CCR}
  \left[\phi(f_1),\phi(f_2)\right]=-i\Delta(f_1,f_2)I,
\end{equation}
where
\begin{equation}
  \label{eq:def-nabla}
  \Delta(f_1,f_2)\equiv \int_\mathcal{M}\epsilon_\mathcal{M} f_1(x)Ef_2(x)
\end{equation}
for all $f_1,f_2\in C^\infty_0(\mathcal{M})$. As shown in~\cite{wald-1994}, Eq.~(\ref{eq:smeared-quantum-field-definition}) can be obtained by formally integrating Eq.~(\ref{eq:unsmeared-field-operator}) weighed by the test function $f$, i.e.,
\begin{equation}
    \label{eq:smeared-phi-with-f}
    \phi(f)=\int_\mathcal{M}\epsilon_\mathcal{M}\, \phi(x)f(x).
\end{equation}

The above construction has the downside that there are infinitely many choices of $\mathcal{H}$ satisfying properties \mbox{\textbf{(i)}-\textbf{(iii)}} listed below Eq.~\eqref{eq:KG-inner-product} and their respective Fock spaces are, in general, unitarily inequivalent. As discussed in~\cite{landulfo-2016}, this issue can be avoided through the algebraic approach to quantum field theory (QFT). For more details, see Refs.~\cite{wald-1994,KM15}. 

In this work, we will focus on a particular class of states: the \textit{quasifree states}, defined as follows. Given a real inner product $\mu:\mathcal{S}\times\mathcal{S}\rightarrow\mathbb{R}$ satisfying
\begin{equation}
   \label{eq:quasifree-state-condition}
   |\sigma(\varphi_1,\varphi_2)|^2\leq4\mu(\varphi_1,\varphi_1) \mu(\varphi_2,\varphi_2),
\end{equation}
for all $\varphi_1,\varphi_2\in\mathcal{S}$, we define a quasifree state $\omega_\mu$ associated with $\mu$ by the relation
\begin{equation}
    \label{eq:quasifree-state-definition}
    \omega_\mu\left[W(Ef)\right]\equiv e^{-\mu(Ef,Ef)/2},
\end{equation}
for all $f\in C^\infty_0(\mathcal{M})$, where the so-called {\em Weyl operators} $W(Ef)$ are defined by
\begin{equation}
    \label{weyl}
    W(Ef)\equiv e^{i\phi(f)}\,,\,\,f\in C^\infty_0(\mathcal{M}).
\end{equation}
The vacuum, n-particle, and thermal states are examples of quasifree states.

\section{The quantum broadcast channel}
\label{sec:broadcast-channel}

A typical broadcast communication scenario involves the transmission of information between one station (sender) and several receivers who will decode the information independently. Let us consider a model in which one observer, Alice, wants to transmit separate information to two other observers, Bob and Charlie, using the quantum field $\phi$ as a broadcast channel. Suppose that the field is initially in some quasifree state $\omega_\mu$\footnote{We note, however, that the results from this section apply to any algebraic state $\omega$ which satisfies $\omega\left[W(Ef)\right]\in\mathbb{R}^+$ for all $f\in C^\infty_0(\mathcal{M})$.}. Suppose also that the three observers follow arbitrary trajectories in the curved spacetime and that each one of them possesses a two-level quantum system that may interact with the quantum field at their will. The two-dimensional Hilbert spaces associated with Alice's, Bob's, and Charlie's qubits are denoted by $\mathcal{H}_A$, $\mathcal{H}_B$, and $\mathcal{H}_C$, respectively.

\begin{figure}[hbt]
    \centering 
    \includegraphics[width=.4\textwidth]{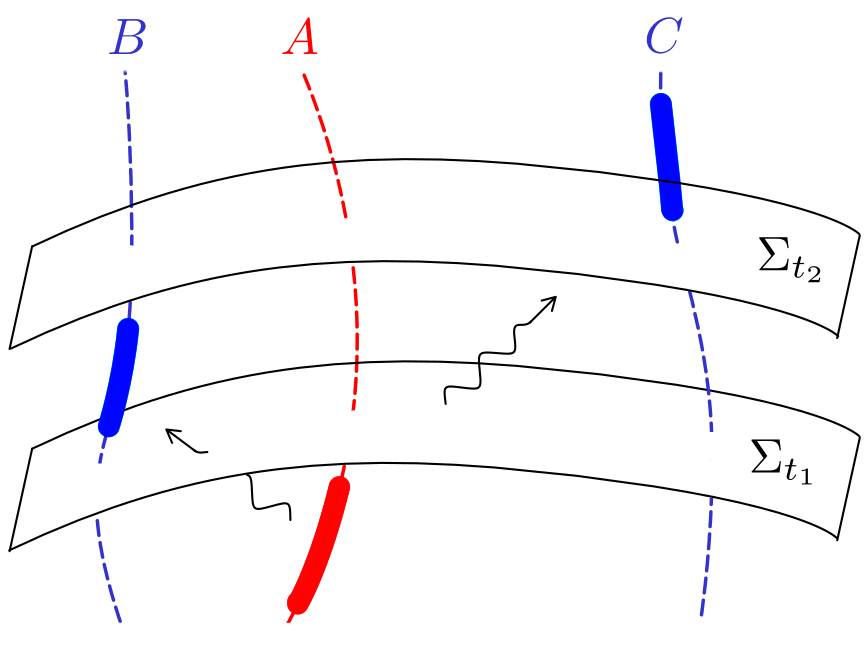}
    \caption{The Figure depicts the quantum broadcast protocol being used.  The dashed lines display the worldlines of the sender, Alice (A, Red), and receivers, Bob and Charlie (B and C, blue). The solid lines in each worldline depict the interaction interval of each observer's qubit with the quantum field. Here, $\Sigma_{t_1}$ and $\Sigma_{t_2}$ represent two Cauchy surfaces of the spacetime.}\label{figQBC}
\end{figure}

The communication setup, illustrated by Fig.~\ref{figQBC}, is as follows: In order to transmit information to Bob and Charlie, Alice prepares her qubit in some initial quantum state $\rho_{-\infty}^A$ and switches on its interaction with the field for a finite time interval $\Delta t_A$ (measured by the parameter $t$). To measure the information imprinted by Alice on the field's state, Bob and Charlie initially prepare their qubits in suitable states~$\rho_{-\infty}^B$ and $\rho_{-\infty}^C$ and then they switch on each of their qubit interaction with the field for finite time intervals $\Delta t_B$ and $\Delta t_C$, respectively. For the sake of simplicity, we will consider here the case where
\begin{itemize}
    \item[\textbf{(QB1)}] Bob lets his qubit interact with the field only after Alice finishes her transmission;
    \item[\textbf{(QB2)}] Charlie lets his qubit interact with the field only after Bob finishes his measurement process.
\end{itemize}

Such communication setup is implemented by means of the Hamiltonian 
\begin{equation}
  \label{eq:total-hamiltonian}
  H(t)\equiv H_\phi(t) + H_{\mathrm{int}}(t),
\end{equation}
where $H_\phi$ is the field Hamiltonian in Eq.~\eqref{eq:field-canonical-hamiltonian} and $H_\mathrm{int}$ is the Hamiltonian that describes the interaction between each qubit and the field which, in the interaction picture, is given by
\begin{equation}
  \label{eq:interaction-hamiltonian}
  H_\mathrm{int}^\mathrm{I} (t)\equiv \sum_{j}\epsilon_j(t)\int_{\Sigma_t}d^3{\bf x} \sqrt{-\mathfrak{g}} \; \psi_j(t,{\bf x}) \phi(t,{\bf x}) \otimes\sigma^{\rm z}_j,
\end{equation}
where $j\in\{A,B,C\}$, with $A$, $B$, and $C$ labeling Alice's, Bob's, and Charlie's qubit, respectively. Here, $\sigma_j^\mathrm{z}$ is one of the Pauli matrices $\left\{\sigma^\mathrm{x}_j,\sigma^\mathrm{y}_j,\sigma_j^\mathrm{z}\right\}$ associated with qubit $j$; $\psi_j(t,\mathbf{x})$ is a smooth real function satisfying $\psi_j|_{\Sigma_t} \in C_0^\infty\left(\Sigma_t\right)$ for all $t$, which models the finite range of interaction between qubit $j$ and the field (i.e., the interaction occurs only at some vicinity of each qubit worldline); and $\epsilon_j(t)$ is a smooth and compactly-supported real {\em coupling function} modeling the finite-time coupling of qubit $j$ with the field. Each coupling function has support
\begin{equation}
  \label{eq:support-coupling-function}
  \mathrm{supp}\;\epsilon_j=\left[T_j^i,T_j^f\right],
\end{equation}
where $T_j^i$ and $T_j^f$ represent the time (with respect to the parameter $t$) in which each qubit interaction with the field is switched-on and -off, respectively. Here, we denote $\Delta t_j\equiv {T_j^f-T_j^i }$. Thus, the hypotheses \textbf{(QB1)} and \textbf{(QB2)} previously listed  can be summarized as
\begin{equation}
    \label{eq:assumption-TC>TB>TA}
    T^i_C\geq T^f_B \geq T^i_B \geq T^f_A.
\end{equation}

The interaction between each qubit and the field given by Eq.~\eqref{eq:interaction-hamiltonian} is very similar to the Unruh-DeWitt model~\cite{dewitt-1979}. However, we assumed that the two levels of each qubit have the same (zero) energy. As we shall see, this assumption allows us to calculate the evolution operator of the system and trace out the field degrees of freedom in a nonperturbative manner, thus making this model interesting to investigate both the causality in the information exchange process as well as the communication between parts lying in early and future asymptotic spacetime regions. We note that one could also have given an energy gap $2\,\delta_j$ for each qubit $j$ in $z$-direction by adding  $H_j=\delta_j\sigma_j^\mathrm{z}$ to the total Hamiltonian in Eq.~\eqref{eq:total-hamiltonian} and still keep the model exactly solvable. This would change it to  
\begin{equation}
  \label{eq:total-hamiltonian-with-gaps}
  H=H_\phi+H_A+H_B+H_C+H_\mathrm{int},
\end{equation}
but would keep the interaction Hamiltonian in the interaction picture, Eq.~\eqref{eq:interaction-hamiltonian}, unchanged. Hence, all the results we will describe below would remain the same.

The interaction-picture time-evolution operator at late times, associated with the foliation $\Sigma_t$, can be written as the time-ordered expression
\begin{equation}
  \label{eq:evolution-operator-definition}
  U\equiv T\exp\left[-i\int_{-\infty}^{\infty}dt\,H^\mathrm{I}_\mathrm{int}(t)\right].
\end{equation}
It can be computed nonperturbatively by using the Magnus expansion~\cite{magnus-expansion}
\begin{equation}
    \label{eq:magnus-expansion}
    U = \exp{\left[\sum_{n=1}^\infty\Omega_n\right]},
\end{equation}
where
\begin{equation}
    \label{eq:omega1-def}
    \Omega_1=-i\int_{-\infty}^\infty dt\, H_\mathrm{int}^\mathrm{I} (t) \;,
\end{equation}
\begin{equation}
    \label{eq:omega2-def}
    \Omega_2=-\frac{1}{2}\int_{-\infty}^\infty dt
    \int_{-\infty}^{t} dt'[H_\mathrm{int}^\mathrm{I}(t)\,,\,H_\mathrm{int}^\mathrm{I} (t')],
\end{equation}
\begin{align}
    \label{eq:omega3-def}
    \!\!\! \Omega_3 = & \frac{i}{6}\int_{-\infty}^\infty \!\!dt \int_{-\infty}^{t} \!\!dt'\int_{-\infty}^{t'} \!\! dt''\{
    [H_\mathrm{int}^\mathrm{I} (t),[H_\mathrm{int}^\mathrm{I} (t'),H_\mathrm{int}^\mathrm{I} (t'')]] \nonumber \\
    & +
    [H_\mathrm{int}^\mathrm{I} (t''),[H_\mathrm{int}^\mathrm{I} (t'),H_\mathrm{int}^\mathrm{I}(t)]]\}, 
\end{align}
and so on. By using Eqs.~\eqref{eq:smeared-phi-with-f}, \eqref{eq:interaction-hamiltonian}, and~\eqref{eq:omega1-def}, we get
\begin{equation}
    \label{eq:omega1}
    \Omega_1=-i\sum_j\phi(f_j)\otimes\sigma_j^\mathrm{z},
\end{equation}
where we have defined
\begin{equation}
    \label{eq:f-def}
    f_j(t,\mathbf{x})\equiv\epsilon_j(t)\psi_j(t,\mathbf{x}).
\end{equation} 
Now, by making use of Eqs.~\eqref{eq:smeared-phi-with-f} and~\eqref{eq:interaction-hamiltonian} together with Eqs.~\eqref{eq:covariant-CCR},~\eqref{eq:assumption-TC>TB>TA}, and~\eqref{eq:omega2-def} we can cast $\Omega_2$ as
\begin{align}
    \label{eq:omega2}
    \Omega_2 = & \,i\Xi I
    -\frac{i}{2}\Delta(f_A,f_B)
    \sigma_A^\mathrm{z}\otimes\sigma_B^\mathrm{z}
    -\frac{i}{2}\Delta(f_A,f_C)
    \sigma_A^\mathrm{z}\otimes\sigma_C^\mathrm{z}
    \nonumber \\
    & 
    -\frac{i}{2}\Delta(f_B,f_C)
    \sigma_B^\mathrm{z}\otimes\sigma_C^\mathrm{z},    
\end{align}
where $\Xi$ is the c-number
\begin{equation*}
  \Xi\equiv\frac{1}{2}\sum_j\int_{-\infty}^\infty dt \; \epsilon_j(t)\int_{-\infty}^{t} \; dt' \epsilon_j(t') \Delta_{j}(t,t'),
\end{equation*}
with
\begin{equation*}
  \Delta_{j}(t,t')\!\equiv \!\!\!\int_{\Sigma_t}\!\!\!\!\!\!d^3{\mathbf{x}}\sqrt{-\mathfrak{g}} \!\!\int_{\Sigma_{t'}}\!\!\!\!\!\!\!d^3{\mathbf{x'}} \sqrt{-\mathfrak{g}'}\psi_j(t,{\mathbf{x}})\Delta(x,x')\psi_j(t',{\mathbf{x'}}),
\end{equation*}
and we recall that $[\phi(x),\phi(x')]\equiv-i\Delta(x,x')I$ is the unsmeared version of Eq.~\eqref{eq:covariant-CCR}. Finally, since $[H_\mathrm{int}^\mathrm{I}(t),H_\mathrm{int}^\mathrm{I} (t')]$ is proportional to the identity, we get
\begin{equation}
    \label{eq:omega3}
    \Omega_k=0\;\;\;\mathrm{for}\; k\geq 3.
\end{equation}
Using the Zassenhaus formula
\begin{equation}
    e^{A+B}=e^Ae^Be^{-\frac{1}{2}[A,B]},
\end{equation}
valid whenever $[A,B]$ is a proportional to the identity, together with Eqs.~\eqref{eq:magnus-expansion},~\eqref{eq:omega1},~\eqref{eq:omega2}, and~\eqref{eq:omega3} we obtain the following unitary evolution operator:
\begin{equation}
    \label{eq:U}
    U = e^{i\Xi}
    e^{-i\phi(f_C)\otimes\sigma_C^\mathrm{z}}
    e^{-i\phi(f_B)\otimes\sigma_B^\mathrm{z}}
    e^{-i\phi(f_A)\otimes\sigma_A^\mathrm{z}}.
\end{equation}

Now that we have the exact evolution operator $U$, we can use it to evolve the initial state of the {3~qubit~+~field} system and then trace out the field and Alice's qubit degrees of freedom. This procedure allows us to obtain the final state of Bob's and Charlie's qubits after the communication protocol has ended. This is the state that they will measure to recover the information that Alice has sent. Explicitly, the final Bob+Charlie state is given by
\begin{equation}
    \label{eq:rhoBC-def}
    \rho^{BC} \equiv \mathrm{tr}_{\phi,A} \left(U
    \rho^A_{-\infty}\otimes\rho^B_{-\infty}\otimes\rho^C_{-\infty}\otimes\rho_{\omega}
    U^\dagger\right),
\end{equation}
where $\rho^j_{-\infty}$ and $\rho_\omega$ are the initial states of qubit $j$ and the field, respectively.

To compute  the trace in Eq.~(\ref{eq:rhoBC-def}), let us cast the operators in Eq.~\eqref{eq:U} as
\begin{equation}
    \label{eq:exp-cast}
    e^{-i\phi(f_j)\otimes\sigma_j^\mathrm{z}} = \cos{[\phi(f_j)]}-i\sin{[\phi(f_j)]}\otimes\sigma_j^\mathrm{z},
\end{equation}
where
\begin{equation}
    \label{eq:cos-def}
    \cos{[\phi(f_j)]}\equiv\frac{1}{2}\left[W(Ef_j)+W(-Ef_j)\right]
\end{equation}
and
\begin{equation}
    \label{eq:sin-def}
    \sin{[\phi(f_j)]}\equiv\frac{1}{2i}\left[W(Ef_j)-W(-Ef_j)\right],
\end{equation}
where $W(Ef)$ is defined in Eq.~(\ref{weyl}). By plugging Eqs.~\eqref{eq:U} and~\eqref{eq:exp-cast} into Eq.~\eqref{eq:rhoBC-def} and then taking the partial traces on $\phi$ and $A$, a direct calculation yields
\begin{align}
    \label{eq:rhoBC-01}
    \rho^{BC}
    & = (\Gamma_{cccccc}+\Gamma_{sccccs})
    \rho^{BC}_{-\infty} \nonumber \\
    & + (\Gamma_{csccsc}+\Gamma_{ssccss})
    \sigma_B^\mathrm{z}
    \rho^{BC}_{-\infty}
    \sigma_B^\mathrm{z} \nonumber \\
    & + (\Gamma_{ccsscc}+\Gamma_{scsscs})
    \sigma_C^\mathrm{z}
    \rho^{BC}_{-\infty}
    \sigma_C^\mathrm{z} \nonumber \\
    & + (\Gamma_{cssssc}+\Gamma_{ssssss})
    \sigma_B^\mathrm{z}\otimes\sigma_C^\mathrm{z}
    \rho^{BC}_{-\infty}
    \sigma_B^\mathrm{z}\otimes\sigma_C^\mathrm{z} \nonumber \\
    & + [(\Gamma_{ccscsc}+\Gamma_{scscss})
    \sigma_B^\mathrm{z}
    \rho^{BC}_{-\infty}
    \sigma_C^\mathrm{z}
    + \mathrm{h.c.}] \nonumber \\
    & - [(\Gamma_{cssccc}+\Gamma_{sssccs})
    \rho^{BC}_{-\infty}
    \sigma_B^\mathrm{z}\otimes\sigma_C^\mathrm{z}
    + \mathrm{h.c.}] \\
    & + [(\Gamma_{cscccs}-\Gamma_{sscccc})
    \langle\sigma_A^\mathrm{z}\rangle_{\rho^A_{-\infty}}
    \rho^{BC}_{-\infty}\sigma_B^\mathrm{z}
    + \mathrm{h.c.}] \nonumber \\
    & + [(\Gamma_{ccsccs}-\Gamma_{scsccc})
    \langle\sigma_A^\mathrm{z}\rangle_{\rho^A_{-\infty}}
    \rho^{BC}_{-\infty}\sigma_C^\mathrm{z}
    + \mathrm{h.c.}] \nonumber \\
    & + [(\Gamma_{csscss}-\Gamma_{ssscsc})
    \langle\sigma_A^\mathrm{z}\rangle_{\rho^A_{-\infty}}
    \sigma_B^\mathrm{z}\rho^{BC}_{-\infty}
    \sigma_B^\mathrm{z}\otimes\sigma_C^\mathrm{z}
    + \mathrm{h.c.}] \nonumber \\
    & + [(\Gamma_{cssscs}-\Gamma_{sssscc})
    \langle\sigma_A^\mathrm{z}\rangle_{\rho^A_{-\infty}}
    \sigma_C^\mathrm{z}\rho^{BC}_{-\infty}
    \sigma_B^\mathrm{z}\otimes\sigma_C^\mathrm{z}
    +\mathrm{h.c.}], \nonumber
\end{align}
where $\mathrm{h.c.}$ stands for Hermitian conjugation, and we have defined
\begin{equation}
    \label{eq:rhoBC-past-def}
    \rho^{BC}_{-\infty}\equiv\rho_{-\infty}^B\otimes\rho_{-\infty}^C,
\end{equation}
\begin{equation}
    \label{eq:<sA>-def}
    \langle\sigma_A^\mathrm{z}\rangle_{\rho^A_{-\infty}}\equiv
    \mathrm{tr}\left(\sigma^\mathrm{z}_A\rho^A_{-\infty}\right),
\end{equation}
and
\begin{align}
    \label{eq:gamma-def}
    \Gamma_{\alpha\beta\gamma\delta\epsilon\zeta}\equiv
    \omega_\mu\big( &
    \mathcal{F}_\alpha[\phi(f_A)]
    \mathcal{F}_\beta[\phi(f_B)]
    \mathcal{F}_\gamma[\phi(f_C)] \nonumber \\
    & \times\mathcal{F}_\delta[\phi(f_C)]\mathcal{F}_\epsilon[\phi(f_B)]\mathcal{F}_\zeta[\phi(f_A)]\big),
\end{align}
with $\alpha,\beta,\gamma,\delta,\epsilon,\zeta\in\{c,s\}$, $\mathcal{F}_c(x)\equiv\cos x,$ and $\mathcal{F}_s(x)\equiv\sin x$. We note that we have written the algebraic field state $\omega_\mu$ as a density matrix with $\mathrm{tr}\left[\rho_\omega W(Ef)\right]\equiv\omega_\mu\left[W(Ef)\right]$.         Furthermore, we have used the fact that the expected value of odd functions of the field operator vanishes since we are assuming that $\omega_\mu$ is a quasifree state (a consequence of Wick's theorem).

Now, each $\Gamma_{\alpha\beta\gamma\delta\epsilon\zeta}$ in Eq.~\eqref{eq:rhoBC-01} can be evaluated by substituting Eqs.~\eqref{eq:cos-def} and~\eqref{eq:sin-def} in Eq.~\eqref{eq:gamma-def} and then using the identity
\begin{equation}
    \label{eq:WEf1WEf2}
    W(Ef_1)W(Ef_2)=e^{\frac{i}{2}\Delta(f_1,f_2)}W[E(f_1+f_2)],
\end{equation}
for all $f,f_1,f_2 \in C^\infty_0(\mathcal{M})$, to simplify the product of the Weyl operators. By substituting these coefficients in Eq.\eqref{eq:rhoBC-01} one finds the explicit form of the state $\rho^{BC}$, which is given in Eq.~\eqref{eq:rhoBC-02} of Appendix~\ref{ap:rhoBC}. The expression in Eq.~\eqref{eq:rhoBC-02} allows one to write the final joint state for Bob's and Charlie's qubits given any initial state configuration for the {3 qubits+field}. 

To define a quantum broadcast channel, we must choose suitable initial states for Bob and Charlie qubits in order to obtain a quantum map relating the initial state of Alice's qubit $\rho^A_{-\infty}$ (which encodes the messages) to the final states that will be probed by them (to decode the messages). Since Bob only performs measurements in his own two-level system, we calculate the expression for the reduced state of his qubit, i.e.,
\begin{equation}
    \rho^B\equiv\mathrm{tr}_C\left(\rho^{BC}\right).
\end{equation}
Taking the trace in Eq.~\eqref{eq:rhoBC-02} relative to Charlie's degrees of freedom, we obtain
\begin{align}
    \label{eq:rhoB-01}
    \rho^B & =\frac{1}{2}\left(1+
    \nu_B\cos{[2\Delta(f_A,f_B)]}\right)
    \rho^{B}_{-\infty} \nonumber \\
    & + \frac{1}{2}\left(1-
    \nu_B\cos{[2\Delta(f_A,f_B)]}\right)
    \sigma_B^\mathrm{z}\rho^{B}_{-\infty}\sigma_B^\mathrm{z} \\
    & + \frac{i}{2}\nu_B\sin{[2\Delta(f_A,f_B)]}
    \langle\sigma_A^\mathrm{z}\rangle_{\rho_{-\infty}^A}
    \left[\rho^{B}_{-\infty},\sigma_B^\mathrm{z}\right],
    \nonumber
\end{align}
where
\begin{equation}
  \label{eq:nu-B-def}
  \nu_B\equiv\omega_\mu\left(W[E(2f_B)]\right)=e^{-2\mu(KEf_B,KEf_B)},
\end{equation}
with $\mu$ be the inner product associated with the field quasifree state $\omega_\mu$ as in Eq.~\eqref{eq:quasifree-state-definition}. Note that it is the last term in Eq.~\eqref{eq:rhoB-01} that contains the information encoded by Alice, and thus it will be useless for Bob to choose the eigenstates $|0\rangle_B$ and $|1\rangle_B$ of $\sigma^\mathrm{z}_B$ as his initial state $\rho^B_{-\infty}$ since this term would vanish. Furthermore, since $\sigma^\mathrm{z}_B$ commutes with the interaction Hamiltonian, he won't recover any information either if he performs projective measurements on this basis. To choose a suitable state $\rho^B_{-\infty}$ that maximizes the chances of success in their communication, suppose for simplicity that Alice encodes a pair of messages in states $\rho^A_{-\infty+}$ and $\rho^A_{-\infty-}$ which will be decoded by Bob using a set of projective measurements in the $x$-direction,
\begin{equation}
    \{ F^B_+\equiv |+\rangle_B{}_B\langle+|,F^B_-\equiv |-\rangle_B{}_B\langle-|\},
\end{equation}
where $\sigma^\mathrm{x}_B|\pm\rangle_B=\pm|\pm\rangle_B$. From Eq.~\eqref{eq:rhoB-01}, we conclude that the probability that Bob measures $l=\pm$ given that Alice has encoded the message $k=\pm$ in $\rho^A_{-\infty k}$ is
\begin{equation}
    \label{eq:probability-lk}
    p(l|k)\equiv\mathrm{tr}\left(F^B_l\rho^{B}_k\right)=\frac{1}{2}(1+l\nu_B\Lambda_k),
\end{equation}
where
\begin{equation*}
    \!\!\!\Lambda_k \!\equiv\!2\mathfrak{R}
    \{
    \beta_B(\cos[2\Delta(f_A,f_B)]
    -i\langle\sigma^\mathrm{z}_A\rangle_{\rho^A_{-\infty k}}\!\!\!
    \sin[2\Delta(f_A,f_B)])\}
\end{equation*}
and $\beta_B\equiv{}_B\langle0|\rho^B_{-\infty}|1\rangle_B$. From these two equations, we see that it is the second term $\Lambda_k$ that contains the information encoded by Alice on her qubit state, and thus we are motivated to choose a state $\rho^B_{-\infty}$ that makes $\beta_B$ a pure imaginary number, which will make the first term of $\Lambda_k$ vanish while maximizing the amplitude of the second term. This motivates us to choose
\begin{equation}
    \label{eq:Bob-initial-state}
    \rho^B_{-\infty}\equiv |y_+\rangle_B{}_B\langle y_+|,
\end{equation}
where
\begin{equation}
    |y_+\rangle_B\equiv
    \frac{1}{\sqrt{2}}\left(|0\rangle_B+i|1\rangle_B\right)
\end{equation}
is an eigenstate of $\sigma^\mathrm{y}_B$ (in this case, $\beta_B=-i/2$). With this choice, we can write Eq.~\eqref{eq:probability-lk} as
\begin{equation}
    p(l|k)=\frac{1}{2}(1-l\nu_B
    \langle\sigma^\mathrm{z}_A\rangle_{\rho^A_{-\infty k}}
    \sin[2\Delta(f_A,f_B)]).
\end{equation}

Now we turn our attention to Charlie. The final reduced state for his qubit is
\begin{equation}
    \rho^C\equiv\mathrm{tr}_B\left(\rho^{BC}\right).
\end{equation}
Taking the trace in Eq.~\eqref{eq:rhoBC-02} relative to Bob's degrees of freedom and using Eq.~\eqref{eq:Bob-initial-state} we obtain
\begin{align}
    \label{eq:rhoC-01}
    \rho^C & =
    \frac{1}{2}(1+
    \nu_C\cos{[2\Delta(f_A,f_C)]}
    \!\cos{[2\Delta(f_B,f_C)]})
    \rho^{C}_{-\infty} \nonumber \\
    + \frac{1}{2} &(1-
    \nu_C\cos{[2\Delta(f_A,f_C)]}
    \!\cos{[2\Delta(f_B,f_C)]})
    \sigma_C^\mathrm{z}\rho^{C}_{-\infty}
    \sigma_C^\mathrm{z} \nonumber \\
    + \frac{i}{2}
    & \nu_C\!\sin{[2\Delta(f_A,f_C)]}
    \!\cos{[2\Delta(f_B,f_C)]}
    \langle\sigma_A^\mathrm{z}\rangle_{\rho_{-\infty}^A}
    \![\rho^{C}_{-\infty},\sigma_C^\mathrm{z}],
\end{align}
where
\begin{equation}
  \label{eq:nu-C-def}
  \nu_C\equiv\omega_\mu\left(W[E(2f_C)]\right)=e^{-2\mu(KEf_C,KEf_C)}.
\end{equation}
To obtain Eq.~\eqref{eq:rhoC-01}, we explicitly used the choice in Eq.~\eqref{eq:Bob-initial-state}, which implies that $\langle\sigma_B^\mathrm{z}\rangle_{\rho^B_{-\infty}}\equiv\mathrm{tr}\left(\sigma^\mathrm{z}_B\rho^B_{-\infty}\right)=0$.
By a completely similar reasoning as the one used to choose Bob's initial state, we are motivated to choose Charlie's initial qubit  state as
\begin{equation}
    \label{eq:Charlie-initial-state}
    \rho^C_{-\infty}\equiv |y_+\rangle_C{}_C\langle y_+|,
\end{equation}
where $\sigma^\mathrm{y}_C|y_+\rangle_C=|y_+\rangle_C$.

Now, the quantum broadcast channel is completely characterized by a linear, completely positive and trace-preserving (CPTP) quantum map $\mathcal{E}$ which takes $\rho^A_{-\infty}$ into a final state $\rho^{BC}$, i.e.,
\begin{equation}
    \rho^{BC}=\mathcal{E}(\rho^A_{-\infty}).
    \label{Edef}
\end{equation}
By substituting the initial states of Bob's and Charlie's qubits given in Eqs.~\eqref{eq:Bob-initial-state} and~\eqref{eq:Charlie-initial-state} into Eq.~\eqref{eq:rhoBC-02}, we find the explicit expression for the quantum broadcast channel $\mathcal{E}$. For the sake of clarity, due to its lengthy expression, we write its explicit form in Eq.~\eqref{eq:QBC-map-final-expression} of  Appendix~\ref{ap:rhoBC}.

For later use, we will denote the reduced channels $\mathcal{E}_B:A\rightarrow B$, $\mathcal{E}_C:A\rightarrow C$ by
\begin{align}
    \label{eq:reduced-B-map-def}
    \mathcal{E}_B(\rho^A_{-\infty})
    & \equiv \mathrm{tr}_C\left[
    \mathcal{E}(\rho^A_{-\infty})\right], \\
    \label{eq:reduced-C-map-def}
    \mathcal{E}_C(\rho^A_{-\infty})
    & \equiv \mathrm{tr}_B\left[
    \mathcal{E}(\rho^A_{-\infty})\right],
\end{align}
respectively. It then follows from Eqs.~\eqref{eq:QBC-map-final-expression}, \eqref{eq:reduced-B-map-def},  and~\eqref{eq:reduced-C-map-def} that they can be explicitly written as 
\begin{align}
    \label{eq:reduced-B-map}
    \mathcal{E}_B(\rho^A_{-\infty})
    & = \frac{1}{2}I_B
    + \frac{\nu_B}{2}\cos{[2\Delta(f_A,f_B)]}
    \sigma^\mathrm{y}_B \nonumber \\
    & - \frac{\nu_B}{2}\sin{[2 \Delta(f_A,f_B)]}
    \langle \sigma_A^\mathrm{z} \rangle_{\rho_{-\infty}^A}
    \sigma^\mathrm{x}_B
\end{align}
and
\begin{align}
    \label{eq:reduced-C-map}
    \mathcal{E}_C(&\rho^A_{-\infty})
    = \frac{1}{2}I_C \nonumber \\
    & + \frac{\nu_C}{2}\cos{[2\Delta(f_A,f_C)]}
    \!\cos{[2\Delta(f_B,f_C)]}\sigma^\mathrm{y}_C \\
    & - \frac{\nu_C}{2}\sin{[2 \Delta(f_A,f_C)]}
    \!\cos{[2\Delta(f_B,f_C)]}
    \langle\sigma_A^\mathrm{z}\rangle_{\rho_{-\infty}^A}
    \!\sigma^\mathrm{x}_C. \nonumber
\end{align}
Given an initial state $\rho^A_{-\infty}$ prepared by Alice on her qubit, these expressions for $\mathcal{E}_B$ and $\mathcal{E}_C$ determine the final local states of Bob's and Charlie's qubit, respectively.

\section{Achievable communication rates}
\label{sec:achievable-rates}

Now that we have constructed a model for a relativistic quantum broadcast channel, we can investigate at which rates classical and quantum information can be reliably transmitted by Alice to Bob and Charlie. We first review a few protocols for quantum broadcast communication published in the literature and then we investigate the achievable rates for our quantum broadcast channel $\mathcal{E}$ defined in Eq.~(\ref{Edef}).

\subsection{Unassisted classical communication}
\label{sec:unassisted-classical-communication}

Let us begin with the investigation of unassisted transmission of classical information. We follow the protocol presented in~\cite{SW2015}, where more details can be found. We evaluate achievable rates for our model and then we discuss how causality is explicitly manifest in our results.

Suppose Alice wishes to transmit a common message $m\in M$ intended for both receivers while sending additional personal messages $m_B\in M_B$ and $m_C\in M_C$ intended for Bob and Charlie, respectively. Each message is chosen from one of the following sets,
\begin{equation}
    M=\{1,\cdots,|M|\}\;,\;\;M_j=\{1,\cdots,|M_j|\},
\end{equation}
with $j\in\{B,C\}$ and $|M|$ denoting the cardinality of $M$. Since the broadcast channel $\mathcal{E}$ is noisy, Alice needs to do a suitable block coding on the possible messages and then make $n$ independent uses of the channel in order to be able to reliably convey the information. More precisely, Alice maps each message triple $(m_B,m,m_C)$ to a codeword $x^n(m_B,m,m_C)$ which is then associated with a quantum state $\rho_{x^n(m_B,m,m_C)}^{A_n}$ defined in the space $\mathcal{H}_A^{\otimes n}$. Then, she transmits $\rho_{x^n(m_B,m,m_C)}^{A_n}$ by making $n$ independent uses of the channel $\mathcal{E}$. The output of the channel is the state
\begin{equation}
    \rho_{x^n(m_B,m,m_C)}^{B_nC_n}
    \equiv \mathcal{E}^{\otimes n}
    \left(\rho_{x^n(m_B,m,m_C)}^{A_n}\right)
\end{equation}
defined on $\mathcal{H}_B^{\otimes n}\otimes\mathcal{H}_C^{\otimes n}$. To decode the message, Bob chooses a positive-operator valued measure (POVM) $\{F^{B_n}_{m_B,m}\,|\,(m_B,m)\in M_B\times M\}$ which acts on the system $B_n$. Similarly, Charlie chooses a POVM $\{G^{C_n}_{m,m_C}\,|\,(m,m_C)\in M\times M_C\}$ which acts on the system $C_n$. We say that an error has occurred when at least one message is incorrectly decoded. Hence, the error probability associated with the transmission of the triple $(m_B,m,m_C)$ is
\begin{equation*}
    p_e(m_B,m,m_C)\equiv
    1-\mathrm{tr}\left[
    \!(F^{B_n}_{m_B,m}\otimes G^{C_n}_{m,m_C})\rho_{x^n(m_B,m,m_C)}^{B_nC_n}
    \!\right].
\end{equation*}

The transmission rates associated with each message are defined as
\begin{equation}
    R\equiv \frac{1}{n} \log_2|M|\;,\;\;
    R_j\equiv \frac{1}{n} \log_2|M_j|.
\end{equation}
These rates essentially measure how many bits of classical information are sent per channel use. If, given an $\epsilon>0$, the average probability of error $\overline{p}_e$ is bounded by $\epsilon$, i.e.,
\begin{equation}
    \overline{p}_e\equiv\frac{1}{|M_B||M||M_C|}\sum_{m_B,m,m_C}p_e(m_B,m,m_C)\leq \epsilon,
\end{equation} 
the \textit{classical-quantum} broadcast channel coding protocol described above is said to be a $(n,R_B,R,R_C,\epsilon)$ code.  We say that a rate triple $(R_B,R,R_C)$ is achievable if given $\epsilon,\delta>0$ there exists a $(n,R_B-\delta,R-\delta,R_C-\delta,\epsilon)$ code for sufficiently large $n$. Hence, saying that a rate triple is achievable means that classical information can be reliably transmitted at rates arbitrarily close to them. 

The achievable rates depend highly on the coding and decoding techniques chosen by the sender and receivers. The best known achievable rate region for general broadcast channels is attained through the so-called \textit{Marton coding scheme}. Following~\cite{SW2015}, we investigate here the quantum version of this protocol.

Suppose for simplicity that no common message is meant to be sent, i.e., let us consider a $(R_B,0,R_C)$ quantum broadcast channel. In this scenario, one strategy they can use is the \textit{Marton coding scheme}, where one chooses two correlated random variables $U$ and $V$, with joint probability distribution denoted by $p$ and reduced probability distributions denoted by $p_U$ and $p_V$. Such a pair of random variables is usually referred to as \textit{binning variables}. Then, for each $m_B\in M_B$ and $m_C\in M_C$, one generates codewords $u^n(m_B)$ and $v^n(m_C)$ according to the reduced probability distributions $p_U(u)$ and $p_V(v)$. Next, the codewords are  mixed together into a single codeword $x^n(m_B,m_C)$ according to a deterministic function $x=f(u,v)$. With this approach, it follows that a rate pair $(R_B,R_C)$ is achievable if it satisfies~\cite{SW2015}
\begin{align}
    \label{eq:marton-RB}
    0 \leq R_B & \leq  I(U;B)_\sigma, \\
    \label{eq:marton-RC}
    0 \leq R_C & \leq  I(V;C)_\sigma, \\
    \label{eq:marton-RBRC}
    R_B + R_C & \leq I(U;B)_\sigma + I(V;C)_\sigma - I(U,V)_\sigma,
\end{align}
where 
\begin{equation}
I(X;Y)_{\rho}\equiv S(X)_\rho + S(Y)_\rho -S({XY})_\rho
\end{equation}
is the mutual information of a state $\rho^{XY}$, with $$S(\alpha)_\rho=-{\rm tr}\left(\rho^
\alpha \log \rho^\alpha \right),$$ 
$\alpha=X,Y$, being the von Neumann entropy of $\rho^\alpha$, $\alpha=X,Y$. Here, $\rho^X={\rm tr}_Y \rho^{XY}$ and $\rho^Y={\rm tr}_X \rho^{XY}$. The states $\sigma$ in Eqs.~(\ref{eq:marton-RB})-(\ref{eq:marton-RBRC}) are obtained by suitably (partially) tracing out the degrees of freedom of the density matrix 
\begin{equation}
    \label{eq:sigma-UVBC-def}
    \sigma^{UVBC}\equiv\sum_{u,v} p(u,v)
    |u\rangle\langle u|^U
    \otimes|v\rangle\langle v|^V
    \otimes\mathcal{E}\left(
    \rho^{A}_{f(u,v)}\right),
\end{equation}
with $p(u,v)$ being the joint probability distribution of the random variables $U$ and $V$.

We begin our analysis by deriving bounds for the achievable rates through the Marton coding scheme applied to our relativistic quantum broadcast channel. To evaluate Eq.~\eqref{eq:marton-RB}, we take partial traces relative to $V$ and $C$ in Eq.~\eqref{eq:sigma-UVBC-def}, obtaining
\begin{equation}
    \label{eq:sigma-UB}
    \sigma^{UB}\equiv\sum_{u} p_U(u)|u\rangle\langle u|^U
    \otimes \omega^B_u,
\end{equation}
where we have written $p(u,v)=p_{V|U}(v|u)p_U(u)$, whereas
\begin{equation}
    \label{eq:omega-Bu-def}
    \omega^{B}_u\equiv\sum_v p_{V|U}(v|u)
    \mathcal{E}_B\left(\rho^A_{f(u,v)}\right).
\end{equation}
A state like $\sigma^{UB}$ in Eq.~\eqref{eq:sigma-UB} is called a \textit{classical-quantum state}. For this class of states, a straightforward calculation shows that~\cite{wilde-2013}
\begin{equation}
    \label{eq:I(U;B)-01}
    I(U;B)_\sigma=S\left[
    \sum_{u} p_U(u)\omega^B_u
    \right]
    - \sum_{u} p_U(u)
    S\left[\omega^B_u\right].
\end{equation}

In order to compute $\omega^B_u$ and its von Neumann entropy,  let us decompose the initial state of Alice's qubit in terms of Bloch vectors, i.e.,
\begin{equation}
    \label{eq:rho-A-bloch}
    \rho^A_{f(u,v)} = \frac{1}{2}\left(I_A
    +\mathbf{r}_{f(u,v)}\cdot\boldsymbol{\sigma}_A\right),
\end{equation}
where $\mathbf{r}_{f(u,v)}\equiv({x}_{f(u,v)},{y}_{f(u,v)},{z}_{f(u,v)})$, $I_A$ is the identity in $\mathcal{H}_A$, $\boldsymbol{\sigma}_A\equiv(\sigma^\mathrm{x}_A,\sigma^\mathrm{y}_A,\sigma^\mathrm{z}_A)$, and
$\|\mathbf{r}_{f(u,v)}\|^2=x_{f(u,v)}^2+y_{f(u,v)}^2+z_{f(u,v)}^2\leq 1$. From Eqs.~\eqref{eq:reduced-B-map}, \eqref{eq:omega-Bu-def}, and~\eqref{eq:rho-A-bloch} we get
\begin{align}
    \label{eq:omega-Bu}
    \omega^{B}_u
    & = \frac{1}{2}I_B
    + \frac{\nu_B}{2}\cos{[2\Delta(f_A,f_B)]}
    \sigma^\mathrm{y}_B \nonumber \\
    & - \overline{z}_u
    \frac{\nu_B}{2}
    \sin{[2 \Delta(f_A,f_B)]}
    \sigma^\mathrm{x}_B,
\end{align}
where $\overline{z}_u\equiv \sum_v p_{V|U}(v|u){z}_{f(u,v)}$, and thus we can further write
\begin{align}
    \label{eq:average-omega-Bu}
    \omega^B \equiv\sum_u p_U(u)\omega^{B}_u
    & = \frac{1}{2}I_B
    + \frac{\nu_B}{2}\cos{[2\Delta(f_A,f_B)]}
    \sigma^\mathrm{y}_B \nonumber \\
    & - \overline{z}
    \frac{\nu_B}{2}
    \sin{[2 \Delta(f_A,f_B)]}
    \sigma^\mathrm{x}_B,
\end{align}
where $\overline{z}\equiv \sum_{u,v} p(u,v)z_{f(u,v)}$. 

Now, by using standard diagonalization, we find that $\omega^B_u$ has eigenvalues $p^B_u$ and $1-p^B_u$, where
\begin{equation}
    \label{eq:pBu-def}
    p^B_u \equiv
    \frac{1}{2} + \frac{\nu_B}{2}
    \sqrt{\overline{z}_u^2\sin^2{[2\Delta(f_A,f_B)]}+\cos^2{[2\Delta(f_A,f_B)]}},
\end{equation}
whereas $\omega^B$ has eigenvalues $p^B$ and $1-p^B$, with
\begin{equation}
    \label{eq:pB-def}
    p^B \equiv
    \frac{1}{2} + \frac{\nu_B}{2}
    \sqrt{\overline{z}^2\sin^2{[2\Delta(f_A,f_B)]}+\cos^2{[2\Delta(f_A,f_B)]}}.
\end{equation}
Therefore, we can now write Eq.~\eqref{eq:I(U;B)-01} as
\begin{equation}
    \label{eq:I(U;B)-02}
    I(U;B)_\sigma=
    H\left(p^B\right)
    -\sum_u p_U(u) H\left(p^B_u\right),
\end{equation}
where $H(x)\equiv -x\log_2{x}-(1-x)\log_2{(1-x)}$, $x\in[0,1]$. Following similar steps, we can show that
\begin{equation}
    \label{eq:I(V;C)-02}
    I(V;C)_\sigma=
    H\left(p^C\right)
    -\sum_v p_V(v) H\left(p^C_v\right),
\end{equation}
where
\begin{align}
    \label{eq:pCv-def}
    p^C_v \equiv \frac{1}{2}
    &+\frac{\nu_C}{2}|\cos{[2\Delta(f_B,f_C)]}| \\
    &\times\sqrt{\overline{z}_v^2\sin^2{[2\Delta(f_A,f_C)]}+\cos^2{[2\Delta(f_A,f_C)]}} \nonumber
\end{align}
with $\overline{z}_v\equiv \sum_u p_{U|V}(u|v){z}_{f(u,v)}$,
and
\begin{align}
    \label{eq:pC-def}
    p^C \equiv \frac{1}{2}
    & + \frac{\nu_C}{2}|\cos{[2\Delta(f_B,f_C)]}| \\
    &\times\sqrt{\overline{z}^2\sin^2{[2\Delta(f_A,f_C)]}+\cos^2{[2\Delta(f_A,f_C)]}}. \nonumber
\end{align}

Now, let us note that $H(x)$ is a monotonically decreasing function when $x\geq1/2$. From Eqs.~\eqref{eq:pBu-def} and~\eqref{eq:pB-def}, we have
\begin{equation}
    p^B_u\leq
    \frac{1}{2} + \frac{\nu_B}{2}
\end{equation}
and
\begin{equation}
    p^B\geq \frac{1}{2}
    + \frac{\nu_B}{2}
    |\cos{[2\Delta(f_A,f_B)]}|,
\end{equation}
and thus it follows that
\begin{equation}
    H(p^B_u)\geq H\left(
    \frac{1}{2} + \frac{\nu_B}{2}
    \right)
\end{equation}
and
\begin{equation}
    \label{eq:H(pB)}
    H(p^B)\leq H\left(
    \frac{1}{2} + \frac{\nu_B}{2}\left|\cos{[2\Delta(f_A,f_B)]}\right|\right).
\end{equation}

As a result, from Eq.~\eqref{eq:I(U;B)-02}, we conclude that
\begin{equation}
    \label{eq:I(U;B)-03}
    I(U;B)_\sigma \leq \mathcal{C}(\mathcal{E}_B),
\end{equation}
where
\begin{equation}
    \label{eq:classical-capacity-Bob}
    \mathcal{C}(\mathcal{E}_B)\equiv
    H\left( \frac{1}{2}+\frac{\nu_B}{2}\left|\cos{[2\Delta(f_A,f_B)]}\right|
    \right)
    -H\left(\frac{1}{2}+\frac{\nu_B}{2}\right)
\end{equation}
is the classical capacity of the reduced channel $\mathcal{E}_B$, given in Eq.~\eqref{eq:reduced-B-map}, as shown in~\cite{landulfo-2016}. We note that the upper bound in Eq.~\eqref{eq:I(U;B)-03} can be attained if we choose random variables $U,V=\{0,1\}$ with $p(u,v)=1/4$ for all $u,v$, associated with Bloch vectors $\mathbf{r}_{f(0,0)}=\mathbf{r}_{f(0,1)}=(0,0,+1)$ and $\mathbf{r}_{f(1,0)}=\mathbf{r}_{f(1,1)}=(0,0,-1)$. By using such choices together with Eq.~\eqref{eq:marton-RB}, we conclude that Alice can reliably convey classical information to Bob at rates arbitrarily close to $\mathcal{C}(\mathcal{E}_B)$.

Similarly, we can show from Eqs.~\eqref{eq:I(V;C)-02}-\eqref{eq:pC-def} that
\begin{equation}
    \label{eq:I(V;C)-03}
    I(V;C)_\sigma \leq \mathcal{C}(\mathcal{E}_C),
\end{equation}
where
\begin{align}
    \label{eq:classical-capacity-Charlie}
    \mathcal{C}(\mathcal{E}_C) & \equiv H\left( \frac{1}{2}+\frac{\nu_C}{2}
    \left|\cos{[2\Delta(f_B,f_C)]}
    \cos{[2\Delta(f_A,f_C)]}\right|
    \right) \nonumber \\
    & - H\left(\frac{1}{2}+\frac{\nu_C}{2}
    |\cos{[2\Delta(f_B,f_C)]}|\right),
\end{align}
is the classical capacity of the reduced channel $\mathcal{E}_C$ given in Eq.~\eqref{eq:reduced-C-map}. The upper bound can be attained, e.g., if we choose random variables $U,V=\{0,1\}$ with $p(u,v)=1/4$ for all $u,v$, associated with Bloch vectors $\mathbf{r}_{f(0,0)}=\mathbf{r}_{f(1,0)}=(0,0,+1)$ and $\mathbf{r}_{f(0,1)}=\mathbf{r}_{f(1,1)}=(0,0,-1)$. Hence, from Eq.~\eqref{eq:marton-RC}, we conclude that Alice can reliably convey classical information to Charlie as well at rates arbitrarily close to $\mathcal{C}(\mathcal{E}_C)$.

It is important to highlight that causality is explicitly manifest on the bounds of the achievable rates. First, we note that the achievable rates $R_B$ between Alice and Bob are bounded by $\mathcal{C}(\mathcal{E}_B)$, which does not depend on the interaction between Charlie's qubit and the quantum field. This should indeed be the case as, from hypothesis \textbf{(QB2)} in Sec.~\ref{sec:broadcast-channel}, Charlie cannot influence the communication between Alice and Bob since he does not perform any actions before Bob finishes his measurement process. Furthermore,  the presence of the commutator $\Delta(f_B,f_C)$ in Eq.~\eqref{eq:classical-capacity-Charlie} indicates that when Bob and Charlie let their qubits interact with the quantum field in causally connected regions of the spacetime, noise from Bob's actions can influence on the rate $R_C$ of communication between Alice and Charlie. Additionally, we note that whenever $\Delta(f_A,f_j)=0$, we have
\begin{equation}
    \mathcal{C}(\mathcal{E}_j)=0
\end{equation}
for $j=B,C$. Hence, when Alice and Bob (or Charlie) interact with the field in causally disconnected regions of the spacetime, the achievable rate in Eq.~\eqref{eq:marton-RB} (or Eq.~\eqref{eq:marton-RC}) will reduce to $R_B=0$ (or $R_C=0$). 

To this day, no one has been able to prove that the Marton rate region given by Eqs.~\eqref{eq:marton-RB}-\eqref{eq:marton-RBRC} is optimal for general broadcast channels, not even in the classical case. However, it is generally conjectured that the Marton rate region indeed represents the full capacity region of general broadcast channels. If this is the case, then our analysis shows that causality will not be violated when transmitting classical information, no matter which communication protocol is chosen.

\subsection{Unassisted and entanglement-assisted quantum communication}
\label{sec:quantum-communication}

Let us now turn our attention to the communication of quantum information. Following~\cite{DHL2010}, we present a father protocol for entanglement-assisted quantum communication through quantum broadcast channels that can be used to investigate at which rates Alice can send classical or quantum information to Bob and Charlie when they share an unlimited supply of entanglement. This protocol can also be adapted to investigate communication rates for quantum information transmission with no prior shared entanglement. After reviewing the father protocol, we investigate both quantum communication scenarios applied to our model of quantum broadcast channel constructed in Sec.~\ref{sec:broadcast-channel}.

Let us suppose that Alice has access to two quantum systems $T_A$ and $T_{A'}$ while Bob and Charlie possess similar quantum systems $T_B$ and $T_C$, respectively. All systems possesses the same dimension $d_{T_A}\equiv \mathrm{dim}\mathcal{H}_{T_A}$. Suppose further that Alice shares maximally entangled states with both  Bob and Charlie: 

\begin{equation}
    \left|\Phi^{T_AT_k}\right\rangle =
    \frac{1}{\sqrt{d_{T_A}}}\sum_{i=0}^{d_{T_A}-1}|i\rangle_{T_A}\otimes|i\rangle_{T_k},
\end{equation}
where the above state is defined on $\mathcal{H}_{T_A}\otimes\mathcal{H}_{T_k}$, with $k=B,C$ and $\{|i\rangle_{T_\alpha}\}$ is an orthonormal set of vectors on $\mathcal{H}_{T_\alpha}$, $\alpha=A,B,C$.

In order to study the transmission of quantum information, we first note that whenever Alice is able to transmit the entanglement she shares with some reference system to each receiver, she will be able to send arbitrary quantum states to each of them. Hence, suppose that Alice possesses two quantum systems $A_1$ and $A_2$ respectively entangled with reference systems $R_1$ and $R_2$ and that these systems are in states $|\Phi^{A_jR_j}\rangle$ defined on $\mathcal{H}_{A_j}\otimes\mathcal{H}_{R_j}$ for $j=1,2$\footnote{As a result, the quantum state being transmitted by Alice to the receiver $j$, $j=B,C$, is $\rho^{A_j}\equiv {\rm tr}_{R_j}|\Phi^{A_jR_j}\rangle \langle \Phi^{A_jR_j}|.$}. Her goal is to send her share of $|\Phi^{A_1R_1}\rangle$ and $|\Phi^{A_2R_2}\rangle$ to Bob and Charlie, respectively. 

The initial global state of the system is
\begin{equation}
    \left|\varphi\right\rangle \equiv
    \left|\Phi^{A_1R_1}\right\rangle\left|\Phi^{A_2R_2}\right\rangle\left|\Phi^{T_AT_B}\right\rangle\left|\Phi^{T_{A'}T_C}\right\rangle
\end{equation}
and we will denote $\rho_\varphi \equiv \left|\varphi\right\rangle\left\langle\varphi\right|$. In order to use the quantum channel $\mathcal{E}$ to share her entanglement with $R_1$ and $R_2$ to Bob and Charlie (and hence, convey quantum information), Alice uses a CPTP map $\mathcal{C}:{\mathcal{H}_{A_1}\otimes \mathcal{H}_{A_2}\otimes \mathcal{H}_{T_A}\otimes \mathcal{H}_{T_{A'}}}\rightarrow \mathcal{H}_A^{\otimes n}$ in order to encode her shares of the quantum systems ($T_A, T_{A'}, A_1,$ and $A_2$), into a state of $n$ qubits. The global state will then reads
\begin{equation}
    \tilde{\rho}^{A_nR_1R_2T_BT_C}\equiv
    \left(
    \mathcal{C}
    \otimes
    I^{R_1R_2T_BT_C}\right)
    \left(\rho_\varphi\right),
\end{equation}
where $I^{R_1R_2T_BT_C}$ is the identity operator of the joint system ${R_1R_2T_BT_C}$. Next, by making $n$ independent uses of the channel $\mathcal{E}$, Alice  sends her total encoded state to Bob and Charlie, which results in the global state
\begin{equation}
    \omega^{B_nC_nR_1R_2T_BT_C} \equiv
    \left(
    \mathcal{E}^{\otimes n}
    \otimes
    I^{R_1R_2T_BT_C}\right)
    \left(\tilde{\rho}^{A_nR_1R_2T_BT_C}\right).
\end{equation}

Bob and Charlie decode their share of the global state by using the CPTP maps $\mathcal{D}_B:{\mathcal{H}_B^{\otimes n}\otimes\mathcal{H}_{T_B}} \rightarrow \mathcal{H}_{B'}$ and $\mathcal{D}_C:{\mathcal{H}_C^{\otimes n}\otimes\mathcal{H}_{T_C}} \rightarrow \mathcal{H}_{C'}$, respectively.  Hence, the final global state is
\begin{equation}
    \zeta^{B'C'R_1R_2}\equiv
    \left(
    \mathcal{D}_C\otimes
    \mathcal{D}_B\otimes
    I^{R_1R_2}\right)
    \left(\omega^{B_nC_nR_1R_2T_BT_C}\right).
\end{equation}

We define the \textit{entanglement-assisted quantum communication rates} as
\begin{equation}
    \widetilde{Q}_B\equiv\frac{1}{n}\log_2 d_{A_1}
    \;,\;\;
    \widetilde{Q}_C\equiv\frac{1}{n}\log_2 d_{A_2},
\end{equation}
where $d_{A_j}\equiv \mathrm{dim}\mathcal{H}_{A_j}$ and $j=1,2$. These rates of quantum communication measure how many qubits are being sent per channel use.

The communication process will be good if given a small $\epsilon>0$ we have
\begin{equation}
    \label{eq:father-protocol-error}
    \left\|
    \zeta^{B'C'R_1R_2} - \rho_\varphi^{B'C'R_1R_2}
    \right\|_1\leq\epsilon,
\end{equation}
where
\begin{equation}
    \left\|\mathcal{O}\right\|_1\equiv
    \mathrm{tr}\left(
    \sqrt{\mathcal{O}^\dagger\mathcal{O}}\right)
\end{equation}
is the trace norm of an operator $\mathcal{O}$. Here, $\rho_\varphi^{B'C'R_1R_2}$ is the analogous of the initial state in the composite system $B'C'R_1R_2$, i.e., given the initial state in Alice's laboratory
\begin{equation}
    \rho_\varphi^{A_1A_2R_1R_2}\equiv
    \left|\Phi^{A_1R_1}\right\rangle
    \left\langle\Phi^{A_1R_1}\right|\otimes
    \left|\Phi^{A_2R_2}\right\rangle
    \left\langle\Phi^{A_2R_2}\right|,
\end{equation}
we define
\begin{equation}
    \rho_\varphi^{B'C'R_1R_2}\equiv
    \left(
    \mathcal{I}^{A_1\rightarrow B'}\otimes
    \mathcal{I}^{A_2\rightarrow C'}\right)
    \left(\rho_\varphi^{A_1A_2R_1R_2}\right),
\end{equation}
where $\mathcal{I}^{A_1\rightarrow B'}$ (or $\mathcal{I}^{A_2\rightarrow C'}$) is the identity map between the quantum systems $A_1$ (or $A_2)$ and $B'$ (or $C'$).

The communication protocol described here is named as a $(n,\widetilde{Q}_B,\widetilde{Q}_C,\epsilon)$ code if it satisfies Eq.~\eqref{eq:father-protocol-error} for every input state $\rho_\varphi^{A_1A_2R_1R_2}$. Again, we say that a rate pair $(\widetilde{Q}_B,\widetilde{Q}_C)$ is achievable if given any $\epsilon,\delta>0$ there exists a $(n,\widetilde{Q}_B-\delta,\widetilde{Q}_C-\delta,\epsilon)$ code for sufficiently large $n$.

Now, given a general broadcast channel $\mathcal{E}:A\rightarrow BC$ and an arbitrary mixed state $\rho^{AA_1A_2}$ defined on $\mathcal{H}_{A}\otimes\mathcal{H}_{A_1}\otimes\mathcal{H}_{A_2}$, it can be shown~\cite{DHL2010} that a entanglement-assisted quantum rate pair $(\widetilde{Q}_B,\widetilde{Q}_C)$ is achievable if
\begin{align}
    \label{eq:EA-QB}
    0\leq\widetilde{Q}_B
    & \leq \frac{1}{2}I(A_1;B)_\sigma, \\
    \label{eq:EA-QC}
    0\leq\widetilde{Q}_C
    & \leq \frac{1}{2}I(A_2;C)_\sigma, \\
    \label{eq:EA-QBQC}
    \widetilde{Q}_B +\widetilde{Q}_C & \leq
    \frac{1}{2}[
    I(A_1;B)_\sigma+I(A_2;C)_\sigma-I(A_1;A_2)_\sigma],
\end{align}
where the mutual information quantities are evaluated relative to the state
\begin{equation}
    \label{eq:sigma-A1A2BC-def}
    \sigma^{A_1A_2BC}\equiv \left(
    \mathcal{E}\otimes I^{A_1A_2}\right)
    \left(\rho^{AA_1A_2}\right).
\end{equation}

In addition to entanglement-assisted quantum communication, the father protocol presented here can be adapted to obtain achievable rates for unassisted quantum communication. To this end, we simply ignore the existence of the quantum systems $T_A,T_{A'},T_B$, and $T_C$ and follow the exact same procedure. As shown in~\cite{DHL2010}, given an arbitrary mixed state $\rho^{AA_1A_2}$ defined on $\mathcal{H}_{A}\otimes\mathcal{H}_{A_1}\otimes\mathcal{H}_{A_2}$, it follows that the following unassisted quantum rate region is achievable:
\begin{align}
    \label{eq:unassisted-QB}
    0\leq Q_B
    & \leq I(A_1\rangle B)_\sigma, \\
    \label{eq:unassisted-QC}
    0\leq Q_C
    & \leq I(A_2\rangle C)_\sigma,
\end{align}
where $\sigma$ is given by Eq.~\eqref{eq:sigma-A1A2BC-def} and $$I(A\rangle B)_\sigma\equiv S(B)_\sigma - S(AB)_\sigma$$ is the quantum coherent information between systems $A$ and $B$.

Now, let us return to the relativistic quantum broadcast channel constructed in Sec.~\ref{sec:broadcast-channel}. To analyze if Alice can send entanglement (and, as a result, an arbitrary state $\rho^A$) to Bob through the broadcast channel, let us note that we may purify the mixed state $\rho^{AA_1A_2}$ by adding an environment system $E$ such that
\begin{equation}
    \label{eq:purifying-state}
    \rho^{AA_1A_2}=\mathrm{tr}_E\left(
    |\psi^{AA_1A_2E}\rangle
    \langle\psi^{AA_1A_2E}|\right),
\end{equation}
where $|\psi^{AA_1A_2E}\rangle\in{\mathcal{H}_{A}\otimes\mathcal{H}_{A_1}\otimes\mathcal{H}_{A_2}\otimes\mathcal{H}_{E}}$ is a pure state. Let us decompose it as
\begin{equation}
    \label{eq:purified-initial-state}
    |\psi^{AA_1A_2E}\rangle =
    \sum_{a=0}^1
    \sum_{a_1=0}^1
    \sum_{a_2=0}^1
    \sum_{e=0}^{d-1} c_{aa_1a_2e}
    |a\rangle_{A}|a_1\rangle_{A_1}
    |a_2\rangle_{A_2}|e\rangle_{E},
\end{equation}
where $|a\rangle_{A}$,$|a_1\rangle_{A_1}$, and $|a_2\rangle_{A_2}$ are  eigenstates of $\sigma^\mathrm{z}_{A}$, $\sigma^\mathrm{z}_{A_1}$, and $\sigma^\mathrm{z}_{A_2}$, respectively. Furthermore, $\{|e\rangle_E\}$ is some orthonormal basis for $\mathcal{H}_E$, with $d\equiv {\rm dim}\mathcal{H}_E$ being as large as needed, and
\begin{equation}
    \sum_{a=0}^1
    \sum_{a_1=0}^1
    \sum_{a_2=0}^1
    \sum_{e=0}^{d-1}
    |c_{aa_1a_2e}|^2=1.
\end{equation}
By defining
\begin{equation}
    \left|\zeta_a\right\rangle_{A_1A_2E}\equiv
    \sum_{a_1=0}^1
    \sum_{a_2=0}^1
    \sum_{e=0}^{d-1} c_{aa_1a_2e}
    |a_1\rangle_{A_1}|a_2\rangle_{A_2}|e\rangle_{E}
\end{equation}
and
\begin{equation}
    \zeta_{aa'}^{A_1A_2} \equiv
    \mathrm{tr}_E\left(
    \left|\zeta_a\right\rangle_{A_1A_2E}{}_{A_1A_2E}
    \left\langle\zeta_{a'}\right||\right),
\end{equation}
we can write Eq.~\eqref{eq:purifying-state} as
\begin{equation}
    \rho^{AA_1A_2} = \sum_{a=0}^1\sum_{a'=0}^1 
    \zeta_{aa'}^{A_1A_2}\otimes
    |a\rangle_{A}{}_A\langle a'|.
    \label{eq:rho-aa1a2}
\end{equation}
By using Eq.~\eqref{eq:rho-aa1a2} in Eq.~\eqref{eq:sigma-A1A2BC-def} and taking the partial trace over $C$ and $A_2$, we obtain
\begin{equation}
    \label{eq:sigma-A1B}
    \sigma^{A_1B} = 
    \sum_{a=0}^1
    \zeta_{aa}^{A_1}\otimes
    \mathcal{E}_B\left(
    |a\rangle_A{}_A\langle a|\right),
\end{equation}
where $\zeta_{aa}^{A_1}\equiv\mathrm{tr}_{A_2}(\zeta_{aa}^{A_1A_2})$ and we have used the fact that
\begin{equation}
    \mathcal{E}_B\left(|a\rangle_A{}_A\langle a'|\right)=
    \delta_{aa'}
    \mathcal{E}_B\left(|a\rangle_A{}_A\langle a|\right),
\end{equation}
which can be proven by a direct calculation using  Eq.~\eqref{eq:reduced-B-map}. We define now the density matrices
\begin{equation}
    \label{eq:def-SBa}
    \mathfrak{S}^B_a\equiv\mathcal{E}_B\left(|a\rangle_A{}_A\langle a|\right),
\end{equation}
and
\begin{equation}
    \tau^{A_1}_{a}\equiv
    \left\|\zeta_{a}\right\|^{-2}
    \zeta_{aa}^{A_1},
\end{equation}
with $\left\|\zeta_{a}\right\|^{2}\equiv\mathrm{tr}(\zeta_{aa}^{A_1})$. This allows us to rewrite Eq.~\eqref{eq:sigma-A1B} as
\begin{equation}
    \label{eq:sigma-A1B-2}
    \sigma^{A_1B} = 
    \sum_{a=0}^1
    \left\|\zeta_{a}\right\|^{2}
    \tau^{A_1}_{a}
    \otimes
    \mathfrak{S}^B_a,
\end{equation}
and we note that $\mathrm{tr}(\mathfrak{S}^B_a)=\mathrm{tr}(\tau^{A_1}_a)=1$ and
\begin{equation}
    \sum_{a=0}^1 \left\|\zeta_{a}\right\|^{2} =
    \sum_{a=0}^1
    \sum_{a_1=0}^1
    \sum_{a_2=0}^1
    \sum_{e=0}^{d-1}
    |c_{aa_1a_2e}|^2=1.
\end{equation}
Hence, we have shown that $\sigma^{A_1B}$ is a separable state, which implies that the reduced channel from Alice to Bob lies in the class of the \textit{entanglement-breaking channels}. As shown in~\cite{holevo-2008}, the coherent information is non-positive for separable states like $\sigma^{A_1B}$, i.e.,
\begin{equation}
    I(A_1\rangle B)_\sigma\leq 0.
\end{equation}
Following similar steps, one can also show that $I(A_2\rangle C)_\sigma\leq0$. As a result, the unassisted achievable rate region given by Eqs.~\eqref{eq:unassisted-QB} and~\eqref{eq:unassisted-QC} reduces to
\begin{equation}
    Q_B=Q_C=0.
\end{equation}
We do not know, to this day, if the region defined by Eqs.~\eqref{eq:unassisted-QB}-\eqref{eq:unassisted-QC} characterizes the full capacity region for general quantum broadcast channels. If this is the case, our analysis implies that Alice cannot send qubits to the receivers without prior shared entanglement. Since the reduced channels $\mathcal{E}_B$ and $\mathcal{E}_C$ are entanglement-breaking, Alice cannot transmit the needed entanglement to establish quantum communication by using only the quantum broadcast channel $\mathcal{E}$.

On the other hand, we can investigate if this limitation changes when the three observers perform an entanglement-assisted quantum communication protocol like the one in the beginning of this section. In this scenario, we recall that Eqs.~\eqref{eq:EA-QB}-\eqref{eq:EA-QBQC} give an achievable (entanglement-assisted) quantum rate region that we shall analyze now.

First, we show that Alice can indeed send quantum information to Bob when they initially share entanglement as follows. We choose the initial input state to be
\begin{equation}
    \label{eq:suitable-rhoAA1A2}
    \rho^{AA_1A_2} \equiv |\Phi^{AA_1}\rangle
    \langle\Phi^{AA_1}|
    \otimes
    \rho^{A_2},
\end{equation}
where $\rho^{A_2}$ is arbitrary and $|\Phi^{AA_1}\rangle$ is the maximally entangled state
\begin{equation}
    |\Phi^{AA_1}\rangle\equiv
    \frac{1}{\sqrt{2}}\sum_{a=0}^1
    |a\rangle_A|a\rangle_{A_1}.
\end{equation}
For this particular state, Eq.~\eqref{eq:sigma-A1B-2} can be written as
\begin{equation}
    \sigma^{A_1B} = \frac{1}{2}
    \sum_{a=0}^1
    |a\rangle_{A_1}{}_{A_1}\langle a|
    \otimes
    \mathfrak{S}^B_a,
\end{equation}
which is a \textit{cq-state} as the one in Eq.~\eqref{eq:sigma-UB}. Following the same steps that led to Eq.~\eqref{eq:I(U;B)-02}, one can show that
\begin{equation}
    I(A_1;B)_\sigma = \mathcal{C}(\mathcal{E}_B),
\end{equation}
where $\mathcal{C}(\mathcal{E}_B)$ is defined in Eq.~\eqref{eq:classical-capacity-Bob}. On the other hand, for this particular input state, we get
\begin{align}
    \sigma^{A_2C} & = \left[
    \sum_{a=0}^1 \frac{1}{2}
    \mathfrak{S}^C_a
    \right] \otimes \rho^{A_2}, \\
    \sigma^{A_1A_2} & = \left[
    \sum_{a=0}^1 \frac{1}{2}
    |a\rangle_{A_1}{}_{A_1}\langle a|
    \right] \otimes
    \rho^{A_2},
\end{align}
where $\mathfrak{S}^C_a\equiv\mathcal{E}_C(|a\rangle_A {}_A\langle a|)$. Since we get these product states,  it follows that
\begin{equation}
    \label{eq:IA2C=0}
    I(A_2;C)_\sigma =
    I(A_1;A_2)_\sigma = 0.
\end{equation}

In view of Eqs.~\eqref{eq:EA-QB}-\eqref{eq:EA-QBQC}, we conclude that Alice will be able to convey quantum information to Bob at a rate arbitrarily close to 
\begin{equation}
    \widetilde{Q}_B=\mathcal{C}(\mathcal{E}_B)/2
\end{equation} 
when they initially share unlimited amounts of entanglement, provided that they let their qubits interact with the field in causally connected regions of the spacetime. Note that this is in contrast with the unassisted case previously discussed. On the other hand, this particular achievable rate region derived here implies that $\widetilde{Q}_C=0$ in view of Eq.~\eqref{eq:IA2C=0}.

Similarly, by switching $A_1$ by $A_2$ in Eq.~\eqref{eq:suitable-rhoAA1A2}, one can show that Alice will be able to transmit quantum states to Charlie (but not to Bob) at a rate arbitrarily close to 
\begin{equation}
\widetilde{Q}_C=\mathcal{C}(\mathcal{E}_C)/2
\end{equation}
when they communicate assisted by shared entanglement. 

Furthermore, initial tripartite entangled states $\rho^{AA_1A_2}$  will, in general, lead to simultaneously nonvanishing rate pairs provided that sender and receivers interact with the field in causally connected regions of spacetime. For example, by choosing the initial input state to be a pure maximally entangled GHZ state
\begin{equation}
    \rho^{AA_1A}= |GHZ\rangle\langle GHZ|,
\end{equation}
where
\begin{equation}
    |GHZ\rangle \equiv \frac{1}{\sqrt{2}}\sum_{a=0}^1
    |a\rangle_A|a\rangle_{A_1}|a\rangle_{A_2},
\end{equation}
one can show by a similar direct calculation that the following entanglement-assisted quantum rate region is achievable:
\begin{align}
    \widetilde{Q}_B
    & \leq \frac{1}{2}\,\mathcal{C}(\mathcal{E}_B), 
    \label{eq:QB-achievable} \\
    \widetilde{Q}_C
    & \leq \frac{1}{2}\,\mathcal{C}(\mathcal{E}_C), \\
    \widetilde{Q}_B +\widetilde{Q}_C & \leq
    \frac{1}{2}
    \left[\mathcal{C}(\mathcal{E}_B)
    +\mathcal{C}(\mathcal{E}_C)-1\right].
    \label{eq:QBQC-achievable-sum-rate}
\end{align}
In~\cite{BL21}, the authors investigate the classical channel capacity $\mathcal{C}(\mathcal{E}_B)$ of the reduced channel $\mathcal{E}_B$ from Alice to Bob for the case where both observers follow inertial or accelerated worldlines on Minkowski spacetime. In some scenarios, the authors show that one can tune the parameters of the channel to achievable capacities close to $1$. Considering that $\mathcal{C}(\mathcal{E}_C)$ has a similar expression, if we consider the case where $\Delta(f_B,f_C)\approx 0$, we can argue in favor of also tune the channel parameters to make $\mathcal{C}(\mathcal{E}_C)$ close to $1$. Under these assumptions, Eqs.~\eqref{eq:QB-achievable}-\eqref{eq:QBQC-achievable-sum-rate} imply that Alice would be able to simultaneously transmit quantum information to Bob and Charlie when assisted by prior entanglement. Hence, the relativistic quantum broadcast constructed in Sec.~\ref{sec:broadcast-channel} seems to impose no limitation to simultaneous entanglement-assisted communication from Alice to both receivers.

\section{Conclusions}
\label{sec:conclusions}

In this paper, we have built a relativistic quantum broadcast channel by using a bosonic quantum field in a general globally hyperbolic spacetime.  In this context, we have explored relativistic effects on the communication of classical and quantum information in a covariant manner, where the parts conveying the information are moving in arbitrary states of motion with the field being in an arbitrary (quasifree) state. 

To construct the quantum broadcast channel, we have considered that Alice (the sender) prepares some input state $\rho^A_{-\infty}$ for her qubit and switches on its interaction with the field for a finite time. After that, Bob (the first receiver) lets his qubit interact with the field for a finite time interval, thus obtaining a final state possibly containing information encoded by Alice. Similarly, after Bob finishes his measurement, Charlie performs an interaction between his qubit and the quantum field to try to recover information imprinted by Alice in the field state. We were able to trace the field degrees of freedom non-perturbatively and showed that suitable initial states for Bob's and Charlie's qubits can be chosen in order to maximize the signaling between Alice and the receivers. This procedure defines a fully relativistic quantum broadcast channel $\mathcal{E}$.

With this channel, we were able to investigate at which rates Alice can reliably convey classical and quantum information to Bob and Charlie. By considering first a scenario where the three observers do not share prior entanglement, we found that Alice can reliably convey classical information to both Bob and Charlie and at which rates she can perform this task. However, we have shown that the broadcast channel presented here breaks entanglement and thus, Alice cannot convey quantum information to Bob and Charlie following an unassisted strategy. Nevertheless, we have shown that this situation changes when they perform entanglement-assisted quantum communication. In this scenario, we were able to find achievable rates that Alice can achieve when sending qubits to the receivers provided that they initially share entangled states.

We were also able to show that all rates that were analyzed here vanish when the interactions between qubits and field occur in causally disconnected regions, an effect that is manifest in all expressions bounding the classical and quantum rates of communication even with the use of quantum resources like entanglement. Thus, our investigation provides good evidence that causality is not violated throughout the communication process, reinforcing the fundamental principles of relativistic physics.

Our study shows that quantum network information theory in general spacetimes is a rich and promising area of research, shedding light on several aspects of the interplay between quantum information theory and relativity. We believe that this work may provide tools to investigate open problems concerning quantum gravity, in particular, the fate of the information that has fallen in (evaporating) black holes. The preservation of causality observed in our analysis reaffirms the robustness of fundamental physical principles, even in the realm of quantum information theory in curved spacetimes. We hope that following the path we presented here could lead us to unveil fundamental aspects of physics that should be present in a full quantum theory of gravity.

\acknowledgments

I. B. and A. L. were fully and partially supported by S\~ao Paulo Research Foundation under Grants 2018/23355-2 and 2017/15084-6, respectively.

\onecolumngrid
\appendix
\section{Full expression for the quantum broadcast channel map}
\label{ap:rhoBC}

As discussed in Sec.~\ref{sec:broadcast-channel}, each $\Gamma_{\alpha\beta\gamma\delta\epsilon\zeta}$ coefficient defined in Eq.~\eqref{eq:gamma-def} can be evaluated by using Eqs.~\eqref{eq:cos-def} and~\eqref{eq:sin-def} together with the product relation given by Eq.~\eqref{eq:WEf1WEf2}. Then, we substitute these coefficients in Eq.~\eqref{eq:rhoBC-01},
obtaining
\begin{align}
    \label{eq:rhoBC-02}
    \rho^{BC}
    & = \frac{1}{4}(1
    +\nu_B\cos{[2\Delta(f_A,f_B)]}
    +\nu_C\cos{[2\Delta(f_A,f_C)]}\cos{[2\Delta(f_B,f_C)]})
    \rho^{BC}_{-\infty} \nonumber \\
    & + \frac{1}{4}(1
    -\nu_B\cos{[2\Delta(f_A,f_B)]}
    +\nu_C\cos{[2\Delta(f_A,f_C)]}
    \cos{[2\Delta(f_B,f_C)]})
    \sigma_B^\mathrm{z}
    \rho^{BC}_{-\infty}
    \sigma_B^\mathrm{z} \nonumber \\
    & + \frac{1}{4}(1
    +\nu_B\cos{[2\Delta(f_A,f_B)]}
    -\nu_C\cos{[2\Delta(f_A,f_C)]}
    \cos{[2\Delta(f_B,f_C)]})\sigma_C^\mathrm{z}
    \rho^{BC}_{-\infty}
    \sigma_C^\mathrm{z} \nonumber \\
    & + \frac{1}{4}(1
    -\nu_B\cos{[2\Delta(f_A,f_B)]}
    -\nu_C\cos{[2\Delta(f_A,f_C)]}
    \cos{[2\Delta(f_B,f_C)]})
    \sigma_B^\mathrm{z}\otimes\sigma_C^\mathrm{z}
    \rho^{BC}_{-\infty}
    \sigma_B^\mathrm{z}\otimes\sigma_C^\mathrm{z} \nonumber \\
    & + \frac{i\nu_B}{4}\sin{[2\Delta(f_A,f_B)]}
    \langle\sigma_A^\mathrm{z}\rangle_{\rho^A_{-\infty}}
    [\rho^{BC}_{-\infty}
    +\sigma_C^\mathrm{z}\rho^{BC}_{-\infty}\sigma_C^\mathrm{z},
    \sigma_B^\mathrm{z}] \nonumber \\
    & + \frac{i\nu_C}{4}
    \sin{[2\Delta(f_A,f_C)]}\cos{[2\Delta(f_B,f_C)]}
    \langle\sigma_A^\mathrm{z}\rangle_{\rho^A_{-\infty}}
    [\rho^{BC}_{-\infty}
    +\sigma_B^\mathrm{z}\rho^{BC}_{-\infty}\sigma_B^\mathrm{z},\sigma_C^\mathrm{z}] \\
    & + \frac{\Lambda^+_{c}}{8}
    \left(
    \rho^{BC}_{-\infty}
    -\sigma_B^\mathrm{z}
    \rho^{BC}_{-\infty}
    \sigma_B^\mathrm{z}
    -\sigma_C^\mathrm{z}
    \rho^{BC}_{-\infty}
    \sigma_C^\mathrm{z}
    +\sigma_B^\mathrm{z}\otimes\sigma_C^\mathrm{z}
    \rho^{BC}_{-\infty}
    \sigma_B^\mathrm{z}\otimes\sigma_C^\mathrm{z}
    \right) \nonumber \\
    & + \frac{\Lambda^-_{c}}{8}
    \left(
    \{\rho^{BC}_{-\infty},
    \sigma_B^\mathrm{z}\otimes\sigma_C^\mathrm{z}\}
    -\sigma_B^\mathrm{z}
    \rho^{BC}_{-\infty}
    \sigma_C^\mathrm{z}
    -\sigma_C^\mathrm{z}
    \rho^{BC}_{-\infty}
    \sigma_B^\mathrm{z}
    \right) \nonumber \\
    & + \frac{i\Lambda^+_{s}}{8}
    \langle\sigma_A^\mathrm{z}\rangle_{\rho^A_{-\infty}}
    [\rho^{BC}_{-\infty}
    -\sigma_C^\mathrm{z}\rho^{BC}_{-\infty}\sigma_C^\mathrm{z}
    ,\sigma_B^\mathrm{z}]
    + \frac{i\Lambda^-_{s}}{8}
    \langle\sigma_A^\mathrm{z}\rangle_{\rho^A_{-\infty}}
    [\rho^{BC}_{-\infty}
    -\sigma_B^\mathrm{z}\rho^{BC}_{-\infty}\sigma_B^\mathrm{z},
    \sigma_C^\mathrm{z}]
    \nonumber \\
    & + \frac{i\nu_C}{4}
    \cos{[2\Delta(f_A,f_C)]}\sin{[2\Delta(f_B,f_C)]}
    \left([\rho^{BC}_{-\infty},
    \sigma_B^\mathrm{z}\otimes\sigma_C^\mathrm{z}]
    +\sigma_B^\mathrm{z}
    \rho^{BC}_{-\infty}
    \sigma_C^\mathrm{z}
    -\sigma_C^\mathrm{z}
    \rho^{BC}_{-\infty}
    \sigma_B^\mathrm{z}
    \right) \nonumber \\
    & - \frac{\nu_C}{4}
    \sin{[2\Delta(f_A,f_C)]}\sin{[2\Delta(f_B,f_C)]}
    \langle\sigma_A^\mathrm{z}\rangle_{\rho^A_{-\infty}}
    \{\rho^{BC}_{-\infty}
    -\sigma_C^\mathrm{z}\rho^{BC}_{-\infty}\sigma_C^\mathrm{z},\sigma_B^\mathrm{z}\}, \nonumber
\end{align}
where we have defined the following coefficients:
\begin{align}
    \Lambda^\pm_{c} & \equiv
    \nu_{BC}^+\cos{[2\Delta(f_A,f_B+f_C)]}
    \pm\nu_{BC}^-\cos{[2\Delta(f_A,f_B-f_C)]}, \\
    \Lambda^{\pm}_{s} & \equiv
    \nu_{BC}^+\sin{[2\Delta(f_A,f_B+f_C)]}
    \pm\nu_{BC}^-\sin{[2\Delta(f_A,f_B-f_C)]}, \\
    \nu_j & \equiv \omega_\mu\left(W[E(2f_j)]\right), \\
    \nu_{BC}^\pm & \equiv \omega_\mu\left(W[E(2f_B\pm2f_C)]\right).
\end{align}

As discussed in Sec.~\ref{sec:broadcast-channel}, we are motivated to fix the initial states for Bob's and Charlie's qubit as given in Eqs.~\eqref{eq:Bob-initial-state} and~\eqref{eq:Charlie-initial-state}. We write these states in terms of their Bloch vectors, i.e.,
\begin{equation}
    \label{eq:initial-state-bloch}
    \rho^j_{-\infty}=\frac{I_j+\sigma^\mathrm{y}_j}{2},
\end{equation}
where $j=B,C$. By substituting Eq.~\eqref{eq:initial-state-bloch} in Eq.~\eqref{eq:rhoBC-02}, and by using the standard commutation relations of the Pauli matrices, we obtain the following expression describing the quantum broadcast channel map:
\begin{align}
    \label{eq:QBC-map-final-expression}
    \mathcal{E}(\rho^A_{-\infty})
    & = \frac{1}{16}(1
    +\nu_B\cos{[2\Delta(f_A,f_B)]}
    +\nu_C\cos{[2\Delta(f_A,f_C)]}\cos{[2\Delta(f_B,f_C)]})
    \left(I_B+\sigma^\mathrm{y}_B\right)\otimes
    \left(I_C+\sigma^\mathrm{y}_C\right) \nonumber \\
    & + \frac{1}{16}(1
    -\nu_B\cos{[2\Delta(f_A,f_B)]}
    +\nu_C\cos{[2\Delta(f_A,f_C)]}
    \cos{[2\Delta(f_B,f_C)]})
    \left(I_B-\sigma^\mathrm{y}_B\right)\otimes
    \left(I_C+\sigma^\mathrm{y}_C\right) \nonumber \\
    & + \frac{1}{16}(1
    +\nu_B\cos{[2\Delta(f_A,f_B)]}
    -\nu_C\cos{[2\Delta(f_A,f_C)]}
    \cos{[2\Delta(f_B,f_C)]})
    \left(I_B+\sigma^\mathrm{y}_B\right)\otimes
    \left(I_C-\sigma^\mathrm{y}_C\right) \nonumber \\
    & + \frac{1}{16}(1
    -\nu_B\cos{[2\Delta(f_A,f_B)]}
    -\nu_C\cos{[2\Delta(f_A,f_C)]}
    \cos{[2\Delta(f_B,f_C)]})
    \left(I_B-\sigma^\mathrm{y}_B\right)\otimes
    \left(I_C-\sigma^\mathrm{y}_C\right) \\
    & - \frac{\nu_B}{4}\sin{[2\Delta(f_A,f_B)]}
    \langle\sigma_A^\mathrm{z}\rangle_{\rho^A_{-\infty}}
    \left(\sigma_B^\mathrm{x}\otimes I_C\right)
    - \frac{\nu_C}{4}
    \sin{[2\Delta(f_A,f_C)]}\cos{[2\Delta(f_B,f_C)]}
    \langle\sigma_A^\mathrm{z}\rangle_{\rho^A_{-\infty}}
    \left(I_B\otimes\sigma_C^\mathrm{x}\right) \nonumber \\
    & + \frac{\Lambda^+_{c}}{8}
    \sigma_B^\mathrm{y}\otimes\sigma_B^\mathrm{y}
    - \frac{\Lambda^-_{c}}{8}
    \left(\sigma_B^\mathrm{x}\otimes
    \sigma_C^\mathrm{y}\right)
    - \frac{\Lambda^+_{s}}{8}
    \langle\sigma_A^\mathrm{z}\rangle_{\rho^A_{-\infty}}
    \left(\sigma_B^\mathrm{x}\otimes
    \sigma_C^\mathrm{y}\right) 
    - \frac{\Lambda^-_{s}}{8}
    \langle\sigma_A^\mathrm{z}\rangle_{\rho^A_{-\infty}}
    \left(\sigma_B^\mathrm{y}\otimes
    \sigma_C^\mathrm{x}\right) \nonumber \\
    & - \frac{\nu_C}{4}
    \cos{[2\Delta(f_A,f_C)]}\sin{[2\Delta(f_B,f_C)]}
    \left(\sigma_B^\mathrm{z}\otimes
    \sigma_C^\mathrm{x}\right)
    - \frac{\nu_C}{4}
    \sin{[2\Delta(f_A,f_C)]}\sin{[2\Delta(f_B,f_C)]}
    \langle\sigma_A^\mathrm{z}\rangle_{\rho^A_{-\infty}}
    \left(\sigma_B^\mathrm{z}\otimes
    \sigma_C^\mathrm{y}\right). \nonumber
\end{align}
By taking partial traces relative to each qubit, one recovers Eqs.~\eqref{eq:reduced-B-map} and~\eqref{eq:reduced-C-map}.

\twocolumngrid

\end{document}